\documentclass[superscriptaddress,preprintnumbers,amsmath,amssymb,prb,twocolumn]{revtex4-1}
\usepackage{amsmath,xcolor,graphicx,colortbl,multirow}
\usepackage{amssymb,soul,color}
\usepackage[left]{lineno}

\begin{document}

\title{Anharmonic electron-phonon coupling in ultrasoft and locally disordered perovskites}

\author{Marios Zacharias}
\email{zachariasmarios@gmail.com}
\affiliation{Univ Rennes, INSA Rennes, CNRS, Institut FOTON - UMR 6082, F-35000 Rennes, France}
\author{George Volonakis}
\affiliation{Univ Rennes, ENSCR, INSA Rennes, CNRS, ISCR - UMR 6226, F-35000 Rennes, France}
\author{Feliciano Giustino}
\affiliation{ Oden Institute for Computational Engineering and Sciences, The University of Texas at Austin,
Austin, Texas 78712, USA%\\This line break forced with \textbackslash\textbackslash
}%
\affiliation{Department of Physics, The University of Texas at Austin, Austin, Texas 78712, USA}
\author{Jacky Even}
\email{jacky.even@insa-rennes.fr}
\affiliation{Univ Rennes, INSA Rennes, CNRS, Institut FOTON - UMR 6082, F-35000 Rennes, France}

\date{\today}

%\linenumbers

\begin{abstract}

Anharmonicity and local disorder (polymorphism) are ubiquitous in perovskite physics, inducing
various phenomena observed in scattering and spectroscopy experiments.
Several of these phenomena still lack interpretation from
first-principles since, hitherto, no approach is available to account for anharmonicity and
disorder in electron-phonon couplings. Here, relying on the special displacement
method, we develop a unified treatment of both and demonstrate that electron-phonon coupling
is strongly influenced when we employ polymorphous perovskite networks. We
uncover that polymorphism in halide perovskites leads to vibrational dynamics far from the ideal noninteracting 
phonon picture and drives the gradual change in their band gap around phase transition temperatures.
We also clarify that combined band gap corrections arising from disorder, spin-orbit coupling, exchange-correlation
functionals of high accuracy, and electron-phonon coupling are all essential.
Our findings agree with experiments, suggesting that polymorphism is the key to address pending
questions on perovskites' technological applications.

\end{abstract}

\maketitle

\noindent {\bf INTRODUCTION} 

\vspace*{0.06cm}
Oxide perovskites are fascinating materials with extensive applications owing to their 
intrinsic ferroelectric, antiferroelectric, and piezoelectric properties~\cite{Gao2020}. 
Halide perovskites are of high interest due to their 
impressive efficiencies in solar cells~\cite{Kojima2009,Snaith2013},  
and attractive applications in optoelectronics, electrocatalysis, and 
thermoelectrics~\cite{Yin2019,Liu2020,Jiang2022}. 
Our understanding of perovskites' key properties is connected to
deviations of the vibrational dynamics and electron-phonon coupling from the standard picture 
observed in conventional semiconductors~\cite{Brenner2016,Philippe2022}. 
For example, halide perovskites exhibit (i) ultralow thermal conductivities attributed to their low-energy
vibrational densities and peculiar anharmonic characteristics~\cite{Lee2017,Simoncelli2019} and 
(ii) limited carrier mobilities 
which have been discussed in terms of electron-phonon Fr\"ohlich coupling~\cite{Ponce2019}, dipolar scattering 
arising from anharmonic halide motion~\cite{Katan2018}, and polaronic transport~\cite{Ghosh2020}.

\begin{figure*}[t!]
\includegraphics[width=0.82\textwidth]{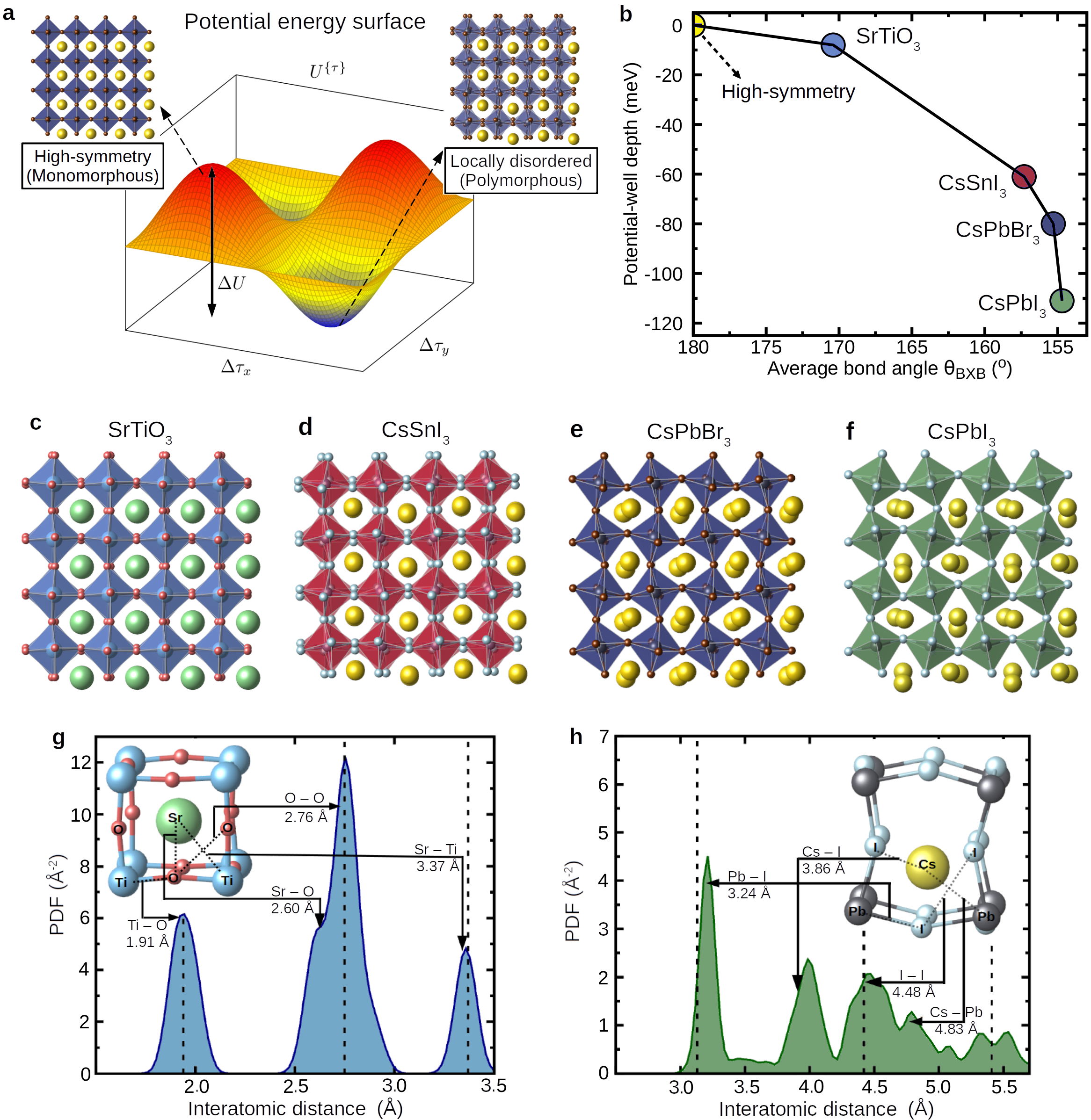}
\caption{ \label{fig1}
{\bf Locally disordered (polymorphous) structures of cubic perovskites.}
{\bf a} Schematic illustration of the potential energy $U$ of cubic perovskites as a function of nuclei displacements
$\Delta \tau$. The local maxima or saddle points for $\Delta \tau = 0$ corresponds to the high-symmetry 
structure with the atoms fixed at their Wyckoff positions. The local minima correspond to locally disordered structures.
{\bf b} Average B-X-B bond angle plotted as a function of the potential-well depth ($\Delta U$) calculated for locally 
disordered cubic SrTiO$_3$, CsSnI$_3$, CsPbBr$_3$, and CsPbI$_3$ using 2$\times$2$\times$2 supercells.
{\bf c-f} Locally disordered structures of SrTiO$_3$ ({\bf c}), CsSnI$_3$ ({\bf d}), 
CsPbBr$_3$ ({\bf e}), and CsPbI$_3$ ({\bf f}). More computational details are available in Methods. 
{\bf g,h} Pair distribution function (PDF) of disordered cubic SrTiO$_3$ ({\bf g}) and CsPbI$_3$ ({\bf h}).
Vertical dashed lines represent pair distribution functions of the high-symmetry structures. 
}
\end{figure*}

A signature of strong anharmonicity in tetragonal or cubic perovskites (stoichiometry ABX$_3$) is the multi-well potential
energy surface (PES), $U$, described by nuclei displacements, $\Delta \tau$, away from 
their static-equilibrium positions (Fig.~\ref{fig1}a). 
Static-equilibrium geometries occur when the net force on each atom vanishes, giving rise to local extrema in the PES. 
The high-symmetry idealized geometry, also referred to as monomorphous
 structure~\cite{Zhao_Zunger2020}, corresponds to a local maximum; 
it features perfectly aligned octahedra and can be described
by a reference unit cell composed of a few atoms. Local minima are explored when the nuclei move away from their
high-symmetry positions, forming a locally disordered (or polymorphous) network 
characterized by tilted BX$_3$ octahedra and a distorted configuration of the A cations (Fig.~\ref{fig1}a).
Description of this form of static or quasi-static (vide infra) disorder requires supercells 
that can accommodate symmetry-breaking domains between the repeated unit cells. 

Typical density functional theory (DFT) calculations of tetragonal or cubic perovskites rely 
on the assumption of a high-symmetry network, disregarding the locally disordered 
ground state configurations. This assumption misses important corrections to the electronic 
structure~\cite{Zhao_Zunger2020,Zhao_Zunger2021} and
requires enforcing the crystal's symmetries on anharmonic phonon dynamics~\cite{Zhao_2021}, thus, 
represented by idealized well-defined dispersions. Such behavior is disconnected from measurements 
of overdamped optical vibrations, structural disorder, and complex pretransitional dynamics close to structural phase 
transitions~\cite{Itoh1994,Even2016,Marronnier2017,Yaffe2017,Marronnier2018,Liu2019,Ferreira2020,Cohen2022,LaniganAtkins2021,Hehlen2022}. 
Furthermore, direct evidence of local disorder in cubic perovskites is observed in 
measurements of pair distribution functions (PDFs), Bragg diffraction, and 
extended diffuse scattering~\cite{Comes1968,Senn2016,Beecher2016,Culbertson2020,Ferreira2020,Doherty2021}. 

\begin{table*} 
\caption{ {\bf Band gaps of high-symmetry and locally disordered perovskites.}
        Relative total energy with respect to high-symmetry structure ($\Delta U$), average B-X-B bond angle 
        ($\bar{\theta}_{\rm BXB}$), band gap ($E_{\rm g}$), and phonon-induced band gap renormalization ($\Delta E_{\rm g}(T)$)
        of high-symmetry (hs) and disordered  (d) cubic SrTiO$_3$, CsPbBr$_3$, CsPbI$_3$, and CsSnI$_3$. 
        Calculations of $E_{\rm g}$ were performed in 2$\times$2$\times$2 supercells using the DFT-PBEsol, 
        HSE, and PBE0 functionals (see Methods). $\Delta E_{\rm g}(T)$ 
        [Eq.~\eqref{eq.ZG_O_T}] was evaluated using ZG displacements in 4$\times$4$\times$4 supercells 
        by taking as a reference the A-SDM IFCs and nuclei positions obtained for 2$\times$2$\times$2~supercells. 
        Temperatures are such that the cubic phase of each compound is thermodynamically stable.
        SOC denotes that spin-orbit coupling is included. Experimental band gaps of cubic SrTiO$_3$, 
        CsPbBr$_3$, CsPbI$_3$, and CsSnI$_3$ are from 
        Refs.~\cite{Kok2015},~\cite{Mannino2020},~\cite{Sutton2018}, 
        and~\cite{Stoumpos2013}, respectively.  \label{table.1}}
        \hspace*{-0.2cm}
%{\footnotesize
% \begin{tabular*}{1.0\textwidth}{l *{8}c }% *{4}c } % columns
% \hline\hline \\  [-0.35 cm]      
%\,\,\,\,\,\,\,\,\,\,\,\,\,\,\,\, &  $\Delta U$ & $\bar{\theta}_{\rm BXB}$ & $E^{\rm DFT}_{\rm g}$ & $E^{\rm DFT}_{\rm g}$ &  $E^{\rm HSE}_{\rm g}$  & $E^{\rm PBE0}_{\rm g}$  & $\Delta E_{\rm g}(T)$  & $E_{\rm g}^{\rm expt.}$  \\ [0.01 cm]
% 
 \begin{tabular*}{1.0\textwidth}{l *{8}c }% *{4}c } % columns
 \hline\hline \\  [-0.35 cm]      
 \,\,\,\,\,\,\,\,\,\,\,\,\,\,\,\,\,\,\,\,\,\,\,\,\,\,\,\, & \,\,\,\,\,\,\, $\Delta U$ \,\,\,\,\,\,\, &\,\,\,\,\,\,\,\, $\bar{\theta}_{\rm BXB}$ \,\,\,\,\,\,\,\, & \,\,\,\,\,\, $E^{\rm DFT}_{\rm g}$ \,\,\,\,\,\, & \,\,\,\,\,\, $E^{\rm DFT}_{\rm g}$ \,\,\,\,\,\, &  \,\,\,\,\,\, \,\,\,\, $E^{\rm HSE}_{\rm g}$ \,\,\,\,\,\,\,\,\,\, & \,\,\,\,\,\, $E^{\rm PBE0}_{\rm g}$ \,\,\,\,\,\,  & \,\,\,\,\,\,  \,\,\,\, $\Delta E_{\rm g}(T)$ \,\,\,\,\,\, \,\,\,\, & \,\,\,\,\,\, $E_{\rm g}^{\rm expt.}$ \,\,\,\,\,\,  \\ [0.01 cm] 
          &  meV/f.u.  & $^\circ$ & eV &  SOC, eV  &  SOC, eV  &  SOC, eV  &   SOC, eV & eV \\ [0.05 cm]  \hline \hline \\ [-0.32 cm] 
 hs-SrTiO$_3$  &  -      &  180.0   & 1.87 & 1.86 & 3.09 & 3.83 &  -0.17 (300K)  & \multirow{2}{*}{3.18} \\ [0.01 cm] 
 d-SrTiO$_3$  &  -8     &  170.4   & 2.11 & 2.12 & 3.40  & 4.15  & -0.25 (300K)  & \\ [0.01 cm] \hline  
 hs-CsPbBr$_3$ &  -      &  180.0   & 1.44 & 0.23 & 1.02 & 1.63 &  0.30 (430K)  & \multirow{2}{*}{2.39} \\ [0.01 cm]
 d-CsPbBr$_3$ &  -80    &  155.3   & 1.94 & 0.80 & 1.68 & 2.30  & 0.09 (430K)   &\\ [0.01 cm] \hline
 hs-CsPbI$_3$  & -       &  180.0   & 1.20 & 0.06 & 0.58 & 1.15  &0.24 (650K)    & \multirow{2}{*}{1.78}   \\ [0.01 cm]
 d-CsPbI$_3$  & -111    &  154.7  & 1.75 & 0.62 & 1.28 & 1.87  & 0.06 (650K)    &\\ [0.01 cm] \hline
 hs-CsSnI$_3$  & -       &  180.0 & 0.12 & -0.27 & 0.0  & 0.0  & 0.23 (500K)    & \multirow{2}{*}{1.3} \\ [0.01 cm]
 d-CsSnI$_3$  & -61     &  157.3 & 0.62 & 0.28  & 0.73  & 1.30  & 0.08 (500K)    &  \\  [0.05 cm] \hline \hline
 \end{tabular*}
%}
\end{table*}

\begin{figure*}[htb!]
 \begin{center}
  \includegraphics[width=0.83\textwidth]{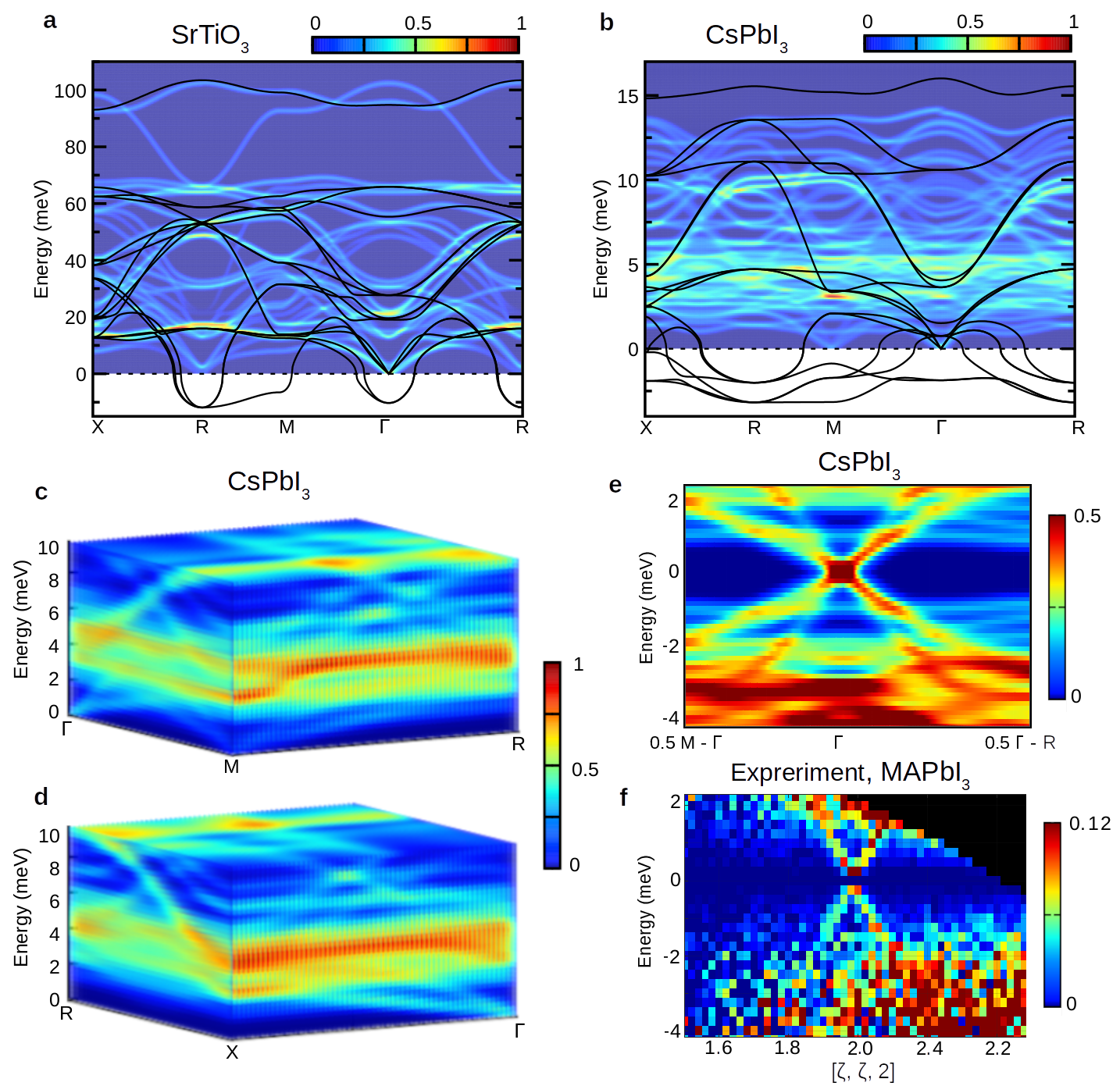}
 \end{center}
\caption{\label{fig2}
{\bf Vibrational spectra at 0~K of locally disordered oxide and halide perovskites.}
{\bf a,b} Phonon spectral functions (color maps) of disordered cubic
SrTiO$_3$ ({\bf a}) and CsPbI$_3$ ({\bf b}) calculated using 2$\times$2$\times$2 supercells
and the phonon unfolding technique (see Methods). Black curves represent phonon dispersions
obtained for the high-symmetry structures.
All calculations include corrections due to long range dipole-dipole interactions. 
{\bf c,d} Phonon spectral function of disordered cubic CsPbI$_3$ visualized in 3D. The spectral function 
spans the momentum plane $\Gamma$-X-R-M. 
{\bf e} Phonon spectral function of disordered cubic CsPbI$_3$ around zone-center. The negative
region, representing loss of energy in neutron scattering experiments, is obtained as a mirror image 
of the positive region.
{\bf f} Experimental data of methylammonium (MA) lead iodide (MAPbI$_3$) measured by 
inelastic neutron scattering (positive or negative energy transfers) at room temperature~\cite{Ferreira2020}.}
\end{figure*}

\begin{figure*}[hbt!]
 \begin{center}
\includegraphics[width=0.8\textwidth]{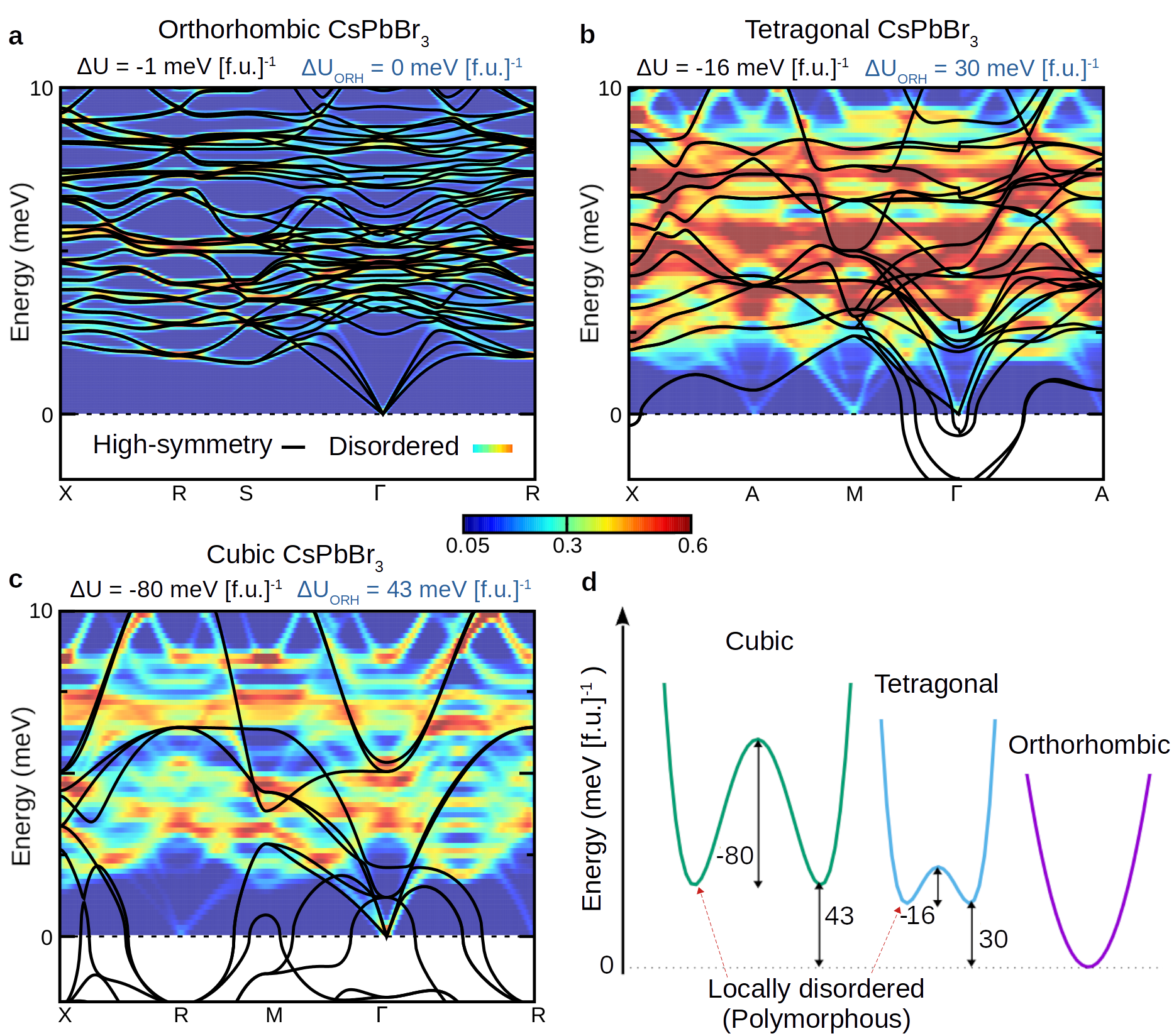}
 \end{center}
\caption{\label{fig3}
{\bf Vibrational spectra at 0~K of locally disordered halide perovskites for different structural phases.}
{\bf a,b,c} Phonon spectral functions (color maps) of disordered orthorhombic ({\bf a}), tetragonal ({\bf b}),
and cubic ({\bf c}) CsPbBr$_3$ calculated using 2$\times$2$\times$2 supercells
and the phonon unfolding technique (see Methods). $\Delta U$ indicates the total energy decrease relative to
the corresponding monomorphous structure, and $\Delta U_{\rm ORH}$ indicates the total energy increase 
with respest to the disordered orthorhombic (ORH) phase. All energies are reported in meV per formula 
unit (f.u., 5 atoms). The harmonic phonon dispersions obtained for the symmetric structures are represented by black curves. 
Corrections due to dipole-dipole interactions are included.
{\bf d} Schematic illustration of the cubic, tetragonal, and orthorhombic PES of CsPbBr$_3$.
}
\end{figure*}

In this work we demonstrate the important role of anharmonicity and local disorder
in the electronic structure, phonon dynamics, and electron-phonon coupling 
of oxide and halide perovskites (SrTiO$_3$, CsPbBr$_3$, CsPbI$_3$, and CsSnI$_3$). 
Hybrid halide perovskites undergo additional relaxations related to molecular reorientations,  
but as a proof of concept we here focus on inorganic compounds. 
We show from first-principles that (i) local disorder and anharmonicity are at the origin of
overdamped and strongly-coupled phonons; 
(ii) local disorder and anharmonicity are essential to describe electron-phonon coupling; 
(iii) low-energy anharmonic optical vibrations dominate thermal band gap renormalization;
(iv) local disorder is the key to explain the smooth evolution of the band gap with temprature
around phase transitions; 
(v) a full description of band gaps and effective masses requires combining disorder with fully relativistic effects.
To address points (i)-(iv) we employ a recently developed approach, namely 
anharmonicity via the special displacement method (A-SDM)~\cite{Zacharias2022}, that 
allows the unified treatment of anharmonic electron-phonon coupling. 
Our study calls for revisiting open questions related 
to electron-phonon and anharmonic properties of halide and oxide perovskites.  

\vspace*{0.2cm}
\noindent {\bf RESULTS}

\noindent {\bf Lattice dynamics} \\
We start the description of lattice dynamics with the harmonic approximation and take 
the expansion of a multi-well PES up to second order in atomic displacements to write:
\begin{eqnarray}  \label{eq.PES_exp}
 U^{\{\tau\}} &=& U_0 +
         \frac{1}{2}  \sum_{i,i'} C_{i, i'} \,  \Delta \tau_{i} \, \Delta \tau_{i'}. 
\end{eqnarray}
$U_0$ is the potential energy with the atoms clamped either at their 
high-symmetry or locally disordered configuration. 
This statistically disordered initial configuration can be obtained using a 
similar procedure (see Methods) to the one described in Ref.~\cite{Zhao_Zunger2020}.
Atomic displacements away from a PES extremum  
are represented by $\Delta \tau_{i}$ where $i$ is a composite index for the
atom, coordinate, and cell. The interatomic force constants (IFCs), defined 
as $C_{i, i'} = \partial^2 U / \partial \tau_{i}  \partial \tau_{i'}$, are used to compute the phonons 
of the high-symmetry or locally disordered structures at 0~K, depending on the stationary point at which 
the second derivatives are evaluated for. 

To incorporate anharmonicity in the lattice dynamics we employ the A-SDM that combines the 
self-consistent phonon theory developed by Hooton~\cite{Hooton1955} and the special displacement method 
(SDM) developed by Zacharias and Giustino (ZG)~\cite{Zacharias_2016,Zacharias_2020}.
In the A-SDM we fixed the nuclei in a supercell at positions determined by ZG displacements
and evaluate the IFCs at temperature $T$ as~\cite{Zacharias2022}:
\begin{eqnarray}  \label{eq.ZG_IFC}
C_{i, i'}(T) \simeq  \frac{\,\, \partial^2 U^{\{\tau^{\rm ZG}\}} } { \partial \tau_{i}  \partial \tau_{i'}}.
\end{eqnarray}
This procedure is performed iteratively until self-consistency in the phonon spectra is achieved. 
The merit of the A-SDM is that the ZG nuclei coordinates, $\{\tau^{\rm ZG}\}$, allow to explore automatically
an effective temperature-dependent harmonic potential that best captures the solution of the nuclear Schr\"odinger equation.
The anharmonic phonons can then be used to describe the crystal's vibrational properties. 
The various schemes used to compute phonon dispersions in this work are described in Supplementary Table~2.

\vspace*{0.1cm}
\noindent {\bf Electron-phonon renormalized observables} \\
Here, we take the A-SDM one step beyond and employ the self-consistent $\{\tau^{\rm ZG}\}$ 
for the nonperturbative evaluation of electron-phonon coupling in anharmonic systems.
Following Ref~\cite{Zacharias_2020}, the renormalization of any temperature-dependent property related to the electronic 
structure can be expressed as: 
\begin{eqnarray}  \label{eq.ZG_O_T}
\Delta O(T) = \frac{1}{2} \sum_{\nu} \frac{\partial^2 O^{\{\tau\}}}{\partial x^2_{\nu}} 
               \sigma^2_{\nu} + \mathcal{O}(x^4_{\nu}),
\end{eqnarray}
where $x_{\nu}$ represents the normal coordinate of the phonon, $\sigma_{\nu}$ is the associated mean-square displacement 
of the atoms, and $\nu$ is a composite index for the band and wavevector.  
The notation $\mathcal{O}(x^4_{\nu})$ represents terms of fourth order and higher in $x_{\nu}$. 
The ZG displacements derived from the A-SDM define the optimum collection of coordinates within
a supercell that allows to compute accurately Eq.~\eqref{eq.ZG_O_T}, describing, at the same time, 
anharmonicity in the PES. In this work, we focus on electron-phonon
renormalized band gaps of tetragonal or cubic perovskites, often described~\cite{Saidi2016} by a harmonic theory 
introduced by Allen and Heine~\cite{Allen_1976}. In this case, the 
derivatives $\partial^2 O^{\{\tau\}}/\partial x^2_{\nu}$ involve linear and second order 
variations of the PES leading to the Fan-Migdal and Debye-Waller self-energy corrections~\cite{Giustino_2017}.
Computing the renormalization starting from the locally disordered structure yields different results 
since (i) electron-phonon self-energy corrections are evaluated for local minima in the PES instead of maxima (c.f. Fig.~\ref{fig1}a),
(ii) electron wavefunctions are modified, and (iii) phonon frequencies are renormalized. 

\begin{figure*}[hbt!]
 \begin{center}
\includegraphics[width=0.86\textwidth]{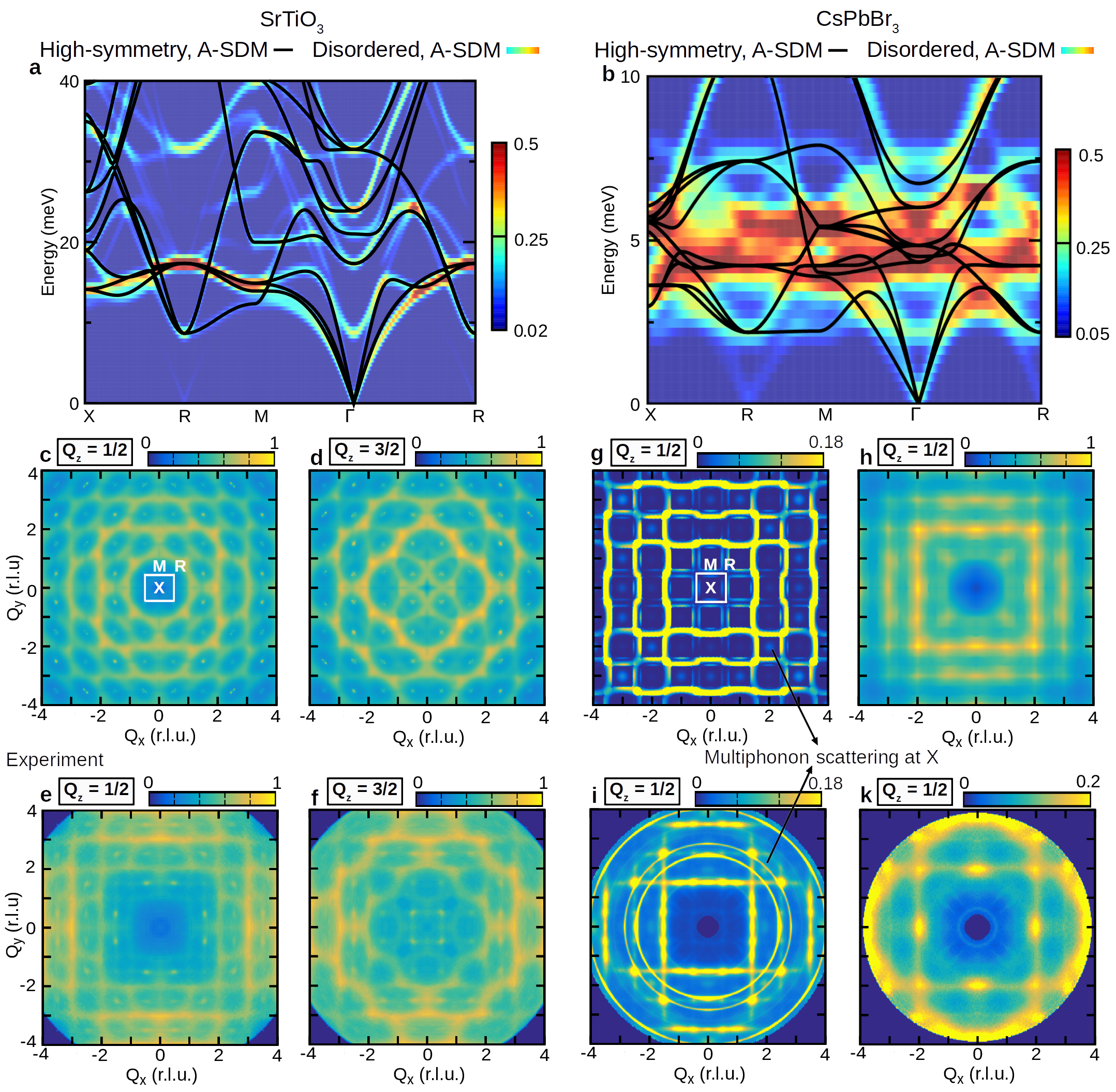}
 \end{center}
\caption{ \label{fig4}
{\bf Temperature-dependent anharmonic phonons of cubic perovskites.}
{\bf a,b} Temperature-dependent anharmonic phonons (black lines) of cubic
SrTiO$_3$ at $300$~K ({\bf a}) and CsPbBr$_3$ at $500$~K ({\bf b}) calculated within the 
A-SDM using 2$\times$2$\times$2 supercells. 
Color maps represent phonon spectral functions of the disordered structures.
{\bf c-f}  Computed and measured diffuse scattering maps at 300~K of cubic SrTiO$_3$ in 
the $(Q_x, Q_y, 0)$ ({\bf c} and {\bf e}) and $(Q_x, Q_y, 1/2)$ ({\bf d} and {\bf f}) reciprocal planes. 
Calculations are performed using the phonons of the disordered network. X-ray scattering data are 
from Ref.~\cite{Kopeck2012}. 
{\bf g-k}  Computed and measured diffuse scattering maps at 500~K of cubic CsPbBr$_3$ in 
 $(Q_x, Q_y, 1/2)$ reciprocal plane. 
Calculations are performed using the A-SDM phonons and high-symmetry network to probe
ultraslow ($<$ 2.5~meV) ({\bf g}) and low-energy (2.5--10~meV) ({\bf h}) phonon dynamics. 
Neutron scattering data are from Ref.~\cite{LaniganAtkins2021} and refer to the energy 
windows of $<$ 2.0~meV ({\bf i}) and 2.0--10~meV ({\bf k}). Diffuse scattering maps are obtained 
within the Laval-Born-James~\cite{Zacharias2021_allph,Zacharias2021_multiph} theory (see Methods).
The scattering wavevector ${\bf Q}$ is expressed in reciprocal lattice units (r.l.u.).
Various schemes used to compute phonons and diffuse scattering are discussed in Supplementary Table~2.
 }
\end{figure*}

\vspace*{-0.06cm}
\noindent {\bf Potential-well depth and relation to local disorder}\\
The depth of the potential-well is, in principle, equivalent to the 
potential energy lowering obtained for the ground state structure 
and provides an indicator of the degree of anharmonicity and static disorder. 
In Table~\ref{table.1} we report the energy lowering ($\Delta U$) and 
average B-X-B bond angle ($\bar{\theta}_{\rm BXB}$) calculated for locally disordered 
cubic SrTiO$_3$, CsSnI$_3$, CsPbBr$_3$, and CsPbI$_3$. 
Our calculations are in good agreement with data reported in Refs.~\cite{Zhao_Zunger2020} 
and~\cite{Zhao_Zunger2021} (see also Supplementary Table~1).
In Fig.~\ref{fig1}b we plot the relationship between local disorder, represented by $\bar{\theta}_{\rm BXB}$,
and the potential-well depth. 
We find that halide perovskites exhibit a considerably higher degree of anharmonicity than SrTiO$_3$ 
which reflects the larger disorder characterizing their ground state 
networks (realized schematically in Figs.~\ref{fig1}c-f) and PDFs (Figs.~\ref{fig1}g,h); 
this finding is connected to the much softer elastic shear modulus~\cite{Piskunov2004,Ferreira2018} 
as well as the different ionicity and bonding nature of halide perovskites~\cite{Piskunov2004,Filip2014}.

\vspace*{0.1cm}
\noindent {\bf Impact of local disorder on phonons at 0~K} \\
In Figs.~\ref{fig2}a,b, we present phonon spectral functions computed 
for disordered SrTiO$_3$ and CsPbI$_3$; for CsSnI$_3$ and CsPbBr$_3$ see Supplementary Figure~1.   
We also include phonon dispersions (black) obtained for high-symmetry structures which display 
large instabilities represented by negative phonon frequencies. 
Importantly, allowing the systems to explore ground state disorder leads to dynamically stable 
phonons (color maps). In Fig.~\ref{fig2}a, we observe band replicas of the $\Gamma$ appearing at the R point and vice-versa;
these features arise from finite size effects and vanish with
the supercell size (Supplementary Figure~2). Local disorder in SrTiO$_3$ also induces a large softening of 
the acoustic branch along R-M with the frequency at R reaching as low as 2~meV.

Remarkably, polymorphism induces extensive broadening and non-dispersive (flattened) optical bands which are overdamped across 
the reciprocal space of CsPbBr$_3$, CsPbI$_3$, and CsSnI$_3$ (Figs.~\ref{fig2}b-d and Supplementary Figure 1). 
Focusing in the frequency region below 4~meV (Fig.~\ref{fig2}e), only the acoustic phonons 
around the $\Gamma$-point are clearly identified. This behavior is consistent with experiments performed 
on lead perovskites~\cite{Fujii1974,Ferreira2020}, suggesting that acoustic phonons emerge from a bath of  
dispersionless optical vibrations (Fig.~\ref{fig2}f). Here, we propose a picture of strongly coupled 
optical vibrations instead of weakly-interacting phonon quasi-particles, since momentum 
information on phonons is smeared out. 
In fact, local disorder, which is distinct from thermal disorder arising
from vibrational fluctuations~\cite{Lahnsteiner2022}, is expected to reduce further the phonon correlation lengths 
and lifetimes of halide perovskites.
Due to its low degree of local disorder, this behavior is not adopted by SrTiO$_3$ which exhibits well-defined phonons 
in the spectral function (Fig.~\ref{fig2}a). The extent of vibrational broadening and coupling is also 
interconnected with the deviation of the PDFs from the archetypal high-symmetry picture (Figs.~\ref{fig1}g,h) and 
lattice softness~\cite{Ferreira2018}. Furthermore, local disorder in halide perovskites causes the decrease 
in energy of optical vibrations, leading to a narrowing of the phonon dispersion and thereby to enhanced 
phonon bunching (Fig.~\ref{fig2}b). 

Figures~\ref{fig3}a,b,c show the phonon spectral functions (color maps) of the three different structural phases of CsPbBr$_3$ calculated using locally disordered networks. In each plot, we report 
the harmonic phonon dispersions of the monomorphous structures (black lines), the potential well-depth ($\Delta U$), and 
the total energy difference with respect to the energy of the orthorhombic phase ($\Delta U_{\rm ORH}$). 
As expected, the level of phonon instabilities in the monomorphous networks are related to the depth of 
$\Delta U$ in each phase. As evidenced by the calculated $\Delta U_{\rm ORH}$, the locally disordered tetragonal and cubic CsPbBr$_3$ 
lie higher in energy than their orthorhombic analog [Fig.~\ref{fig3}d]. It is also apparent that local disorder in the orthorhombic 
structure has a negligible effect on the system's total energy, yielding identical stable phonons with those 
obtained for the monomorphous phase. On the contrary, local disorder relatively to the cubic and tetragonal high-symmetry 
networks is much more prominent, increasing the coupling between individual bands and, hence, suggesting a further
decrease in the lattice thermal conductivity of these phases~\cite{Lee2017}.

\vspace*{0.1cm}
\noindent {\bf Temperature-dependent phonon anharmonicity} \\
Figures~\ref{fig4}a,b show temperature-dependent phonon dispersions of the high-symmetry (black) 
SrTiO$_3$ and CsPbBr$_3$ calculated using the A-SDM~\cite{Zacharias2022}; for CsPbI$_3$ and CsSnI$_3$ see Supplementary Figure~3. 
The phonon spectral functions (color maps) are obtained by combining the atomic positions of locally disordered networks 
with the IFCs obtained by A-SDM (see also discussion around Supplementary Table~2). 
It can be seen for SrTiO$_3$ that the spectral function follows closely the A-SDM phonon 
dispersion. This observation aligns with a picture of non-interacting phonons at 
low temperatures and it is also related with the minimal level of local disorder in SrTiO$_3$ reported 
in Table~\ref{table.1}. On the contrary, symmetry-breaking in halide perovskites induces the coupling of low-energy 
optical vibrations and the reduction of their coherence lengths.
Moreover, accounting for temperature-dependent anharmonicity in our calculations via the A-SDM
reproduces the thermal vibrational softening along R-M (Fig.~\ref{fig4}b and Supplementary Figure~3), 
consistent with previous calculations~\cite{Patrick2015,LaniganAtkins2021}.

\vspace*{0.1cm}
\noindent{\bf Diffuse scattering} \\ 
In Figs.~\ref{fig4}c-h we present thermal diffuse scattering maps of SrTiO$_3$ and CsPbBr$_3$ 
at 300~K and 500~K; for CsPbI$_3$ and CsSnI$_3$ check Supplementary Figure~4.
We find that using the phonons obtained for the disordered networks 
(i.e spectral function in Fig.~\ref{fig2}a) reproduces better the experimental maps~\cite{Kopeck2012} 
of SrTiO$_3$ in $(Q_x, Q_y, 1/2)$ (Fig.~\ref{fig4}e) 
and $(Q_x, Q_y, 3/2)$ (Fig.~\ref{fig4}f) reciprocal planes, where ${\bf Q}= (Q_x, Q_y, Q_z)$ is the scattering wavevector. 
Importantly, in the calculated maps we can identify the emergence of phonon-induced scattering peaks at the R-points
which correspond to the ultrasoft phonons discussed for Fig.~\ref{fig2}a.
These features are present in X-ray diffuse scattering measurements of Ref.~\cite{Kopeck2012} 
and attributed to dynamic disorder due to antiphase rotations of the octahedra, 
mimicked by the static distortions present in our disordered network. 
Note that diffuse scattering at R is absent when the high-symmetry structure with the A-SDM phonons 
calculated for 2$\times$2$\times$2 supercells are combined (Supplementary Figures~5 and~6). 

At variance with SrTiO$_3$, the scattering maps computed for the high-symmetry CsPbBr$_3$ at 500~K
yield better agreement with measurements reported in Ref.~\cite{LaniganAtkins2021}.
To illustrate this we perform calculations of scattering maps in the $(Q_x, Q_y, 1/2)$ plane 
for two separate frequency ranges (Figs.~\ref{fig4}g and~\ref{fig3}h) using the A-SDM phonons.   
Focusing on the scattering induced by ultraslow dynamics ($<$ 2.5~meV), the acoustic
soft branch along R-M leads to the formation of vertical and horizontal diffuse rods 
across several Brillouin zones, as observed in measurements for CsPbBr$_3$ (Fig.~\ref{fig4}i). 
We stress that local disorder induces a hardening of the modes along R-M (Supplementary Figure~1), 
and thus it prevents the formation of diffuse rods (Supplementary Figure~7). This comes as no surprise
since the disordered network should be regarded as a quasi-static approximation that cannot describe 
the ultraslow dynamical octahedral tilting~\cite{LaniganAtkins2021,Hehlen2022} and, thus, the relaxation of the system 
between various (deep) minima in the PES. Focusing on the scattering induced by low-energy excitations (2.5--10~meV), 
we detect broad diffuse rods along M-X (Fig.~\ref{fig4}h) which are 
in close agreement with measurements in the range 2.0--10~meV (Fig.~\ref{fig4}k). 
Now, employing the phonons of the locally disordered network perfectly reproduces the diffuse scattering 
maps (Supplementary Figure~4), demonstrating that, unlike the ultraslow octahedra relaxations ($<$ 2.5~meV), 
the low-energy vibrations (2.5--10~meV) are captured correctly. 

Interestingly, in Fig.~\ref{fig4}g and Supplementary Figure~4, we can identify low intensity multiphonon scattering 
signatures at the X-points arising from the combined momenta of two phonons along M-R. These fine structures are present 
in neutron scattering maps of CsPbBr$_3$ (Fig.~\ref{fig4}i), but not interpreted before. 
Our findings here suggest that low-energy multiphonon excitations are another source of manifestation of 
anharmonicity in halide perovskites, emerging from highly anharmonic zone-edged modes.

\begin{figure*}[t!]
 \begin{center}
\includegraphics[width=0.70\textwidth]{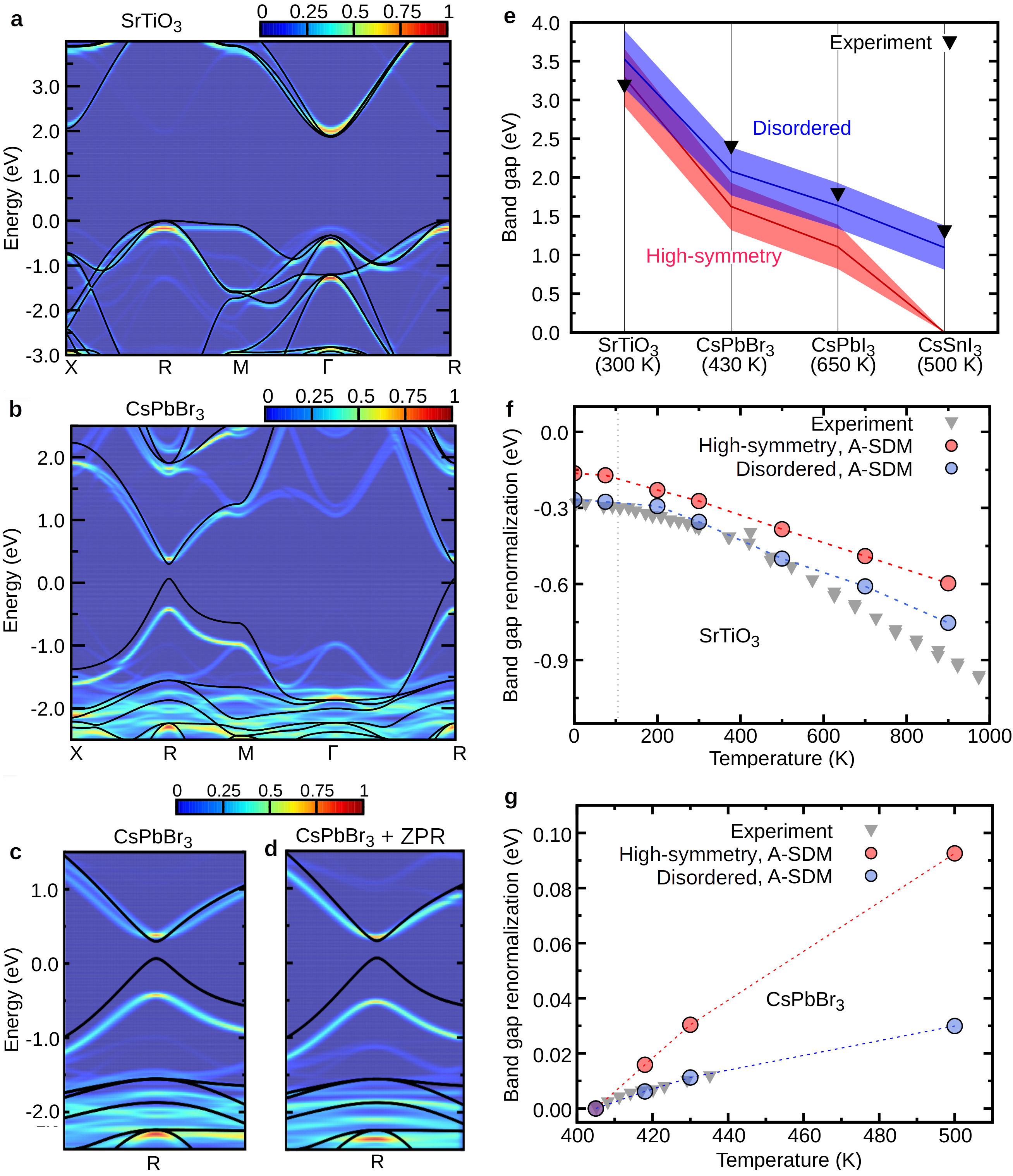}
 \end{center}
\vspace*{-0.4cm}
\caption{\label{fig5}
{\bf Electron-phonon renormalized band gaps of cubic perovskites.}
{\bf a,b} Fully-relativistic electron spectral functions (color maps) of disordered cubic
SrTiO$_3$ ({\bf a}) and CsPbI$_3$ ({\bf b}) calculated using 2$\times$2$\times$2 supercells
and the band unfolding technique~\cite{Zacharias_2020}. Black curves represent the 
band structures of the high-symmetry networks. 
{\bf c} As in {\bf b} but now focusing around the band edges at R. 
{\bf d} As in {\bf c} but now the effect of electron-phonon coupling at 0~K 
[i.e. the ZPR] via the A-SDM is included.
{\bf e} HSE and PBE0 band gaps of high-symmetry and disordered cubic perovskites including 
the effect of SOC and electron-phonon coupling. Lines represent the 
average of HSE and PBE0 band gaps and the shaded areas define the uncertainty.  
Experimental gaps (black) of SrTiO$_3$, CsPbBr$_3$, CsPbI$_3$, and CsSnI$_3$ are from 
Refs.~\cite{Kok2015},~\cite{Mannino2020},~\cite{Sutton2018}, and~\cite{Stoumpos2013}. 
{\bf f,g} Phonon-induced band gap renormalization of high-symmetry (red) and disordered (blue) cubic 
SrTiO$_3$ ({\bf f}) and CsPbBr$_3$ ({\bf g}) as a function of temperature calculated using ZG displacements 
for 6$\times$6$\times$6 and 4$\times$4$\times$4 supercells, respectively. 
The data below the phase transition temperature ($\sim$105~K) of SrTiO$_3$ are 
obtained using the tetragonal structure and ZG displacements for 6$\times$4$\times$4 supercells.
ZG displacements were generated using anharmonic A-SDM IFCs and accounting 
for thermal lattice expansion. For CsPbBr$_3$ the effect of SOC is included and the renormalization is determined 
with respect to $T=405$~K. Optical spectroscopy data (grey) for SrTiO$_3$ and CsPbBr$_3$ are from 
Refs.~\cite{Kok2015} and~\cite{Mannino2020}. Except for ({\bf e}) all calculations are performed within DFT-PBEsol.
}
\end{figure*}

\vspace*{0.1cm}
\noindent {\bf Effect of disorder on anharmonic electron-phonon coupling} \\ 
Figures~\ref{fig5}a,b compare the electron spectral functions
of locally disordered cubic SrTiO$_3$ and CsPbI$_3$ (color maps) with the band structures of their 
high-symmetry counterparts (black); for CsSnI$_3$ and CsPbBr$_3$ check Supplementary Figure~8.
The effect of symmetry-breaking domains on the electronic structure can be understood intuitively by inspection of 
the PDFs (Figs.~\ref{fig1}g,h). It turns out that local disorder induces slight 
changes in the electronic structure of SrTiO$_3$ with the main impact being on the band edges 
at $\Gamma$ and R points (Fig.~\ref{fig5}a). In particular, symmetry lowering induces 
a band gap opening of 0.24 eV which is in agreement with the value reported in Ref.~\cite{Zhao_Zunger2021}. 

Local disorder causes distinct changes on the electronic structure of cubic halide perovskites 
(Fig.~\ref{fig5}b and Supplementary Figure~8). 
Those are large band gap openings, band dispersion renormalization, and band broadening.
The quantitative comparison between the band gaps calculated for the high-symmetry and disordered halide perovskites 
is provided in Table~\ref{table.1}. Our calculations 
reveal a band gap enhancement due to local disorder of more than 0.50~eV for cubic 
CsPbBr$_3$, CsPbI$_3$, and CsSnI$_3$, similarly to previous calculations~\cite{Zhao_Zunger2020}. 
Owing to a higher degree of local disorder, represented by $\bar{\theta}_{\rm BXB}$ in Table~\ref{table.1}, 
halide perovskites exhibit a larger band gap opening than SrTiO$_3$. Interestingly, this observation 
suggests an indirect relationship between anharmonicity and the band gap in perovskite systems. 
In fact, the connection between $\bar{\theta}_{\rm BXB}$ with the band gap is related to the 
changes in the overlap between the metal and halogen states~\cite{Filip2014}.
Moreover, our calculations show that local disorder in the tetragonal and orthorhombic phases 
of CsPbBr$_3$ yields band gap enhancements of 0.17 and 0.001~eV. These values are much lower than the 
one obtained for the cubic phase (0.57~eV), in line with the potential well-depth of each phase (Fig.~\ref{fig3}d).
Our values for the Pb-based compounds in Table~\ref{table.1} show that spin-orbit coupling (SOC) induces a giant gap 
closing of 1.1--1.2~eV, in agreement with Ref.~\cite{Even2013}. 
We find that SOC has also a strong influence on the effective mass enhancement 
due to local disorder (Supplementary Table~3). For instance, excluding SOC, local disorder leads to 
electron and hole mass enhancements $\lambda$ (see Methods) between 0.4--1.2. When SOC is taken into account, 
the disordered networks yield $\lambda$ of 1.3--2.3 for CsPbBr$_3$ and 3.3--4.7 for CsPbI$_3$. We stress that the 
calculated effective masses of disordered CsPbBr$_3$ and CsPbI$_3$, ranging between 0.13--0.20, compare well 
with 0.114 and 0.126 measured from inter-Landau level transitions 
in CsPbI$_3$ and CsPbBr$_3$, respectively~\cite{Yang2017}.
   
The impact of local disorder is clearly manifested in the electronic structure of CsSnI$_3$ 
(Supplementary Figure~8). The fully relativistic band structure of high-symmetry CsSnI$_3$ 
exhibits an artefact in the conduction band minimum, resulting from the exchange of orbital character 
between the band edges. This suggests a metallic-like behavior for CsSnI$_3$ (hence the value -0.27~eV in Table~\ref{table.1})
and leads to unphysical negative electron effective masses at the R-point.  
Instead, accounting for local disorder recovers the standard picture of a parabolic conduction band minimum with positive 
effective masses of 0.08 and a direct gap of 0.28~eV at the R-point. 

The fully relativistic DFT band gaps of disordered perovskites still largely underestimate the experimental 
values~\cite{Kok2015,Mannino2020,Sutton2018,Stoumpos2013}, reported in Table~\ref{table.1}, by more than 1 eV, 
due to the DFT semi-local description of correlation effects. 
As shown in Table~\ref{table.1}, this discrepancy is significantly alleviated when self-energy corrections 
through the HSE and PBE0 hybrid functionals are accounted for. 

In Figs.~\ref{fig5}c,d we compare the electronic structure around the band extrema of disordered CsPbBr$_3$ without and with
the effect of phonon-induced zero-point renormalization (ZPR)~\cite{Zacharias_2020}. Electron-phonon interactions, incorporated by 
ZG displacements in 2$\times$2$\times$2 supercells, induce a band gap opening, yielding a ZPR of 
29~meV. Increasing the supercell size to 4$\times$4$\times$4 reverses the sign of the ZPR and yields
a band gap decrease of 35~meV. We also comment that combining disorder 
with ZG displacements does not lead to an artificial Rashba-Dresselhaus splitting of the doubly degenerate band 
extrema, reflecting that perovskite crystals at thermal equilibrium should maintain centrosymmetricity~\cite{Schlipf2021}. 

Table~\ref{table.1} also reports the phonon-induced band gap renormalization, $\Delta E_g(T)$, of 
cubic SrTiO$_3$, CsPbBr$_3$, CsPbI$_3$, and CsSnI$_3$ at $300$, $430$, $650$, and $500$~K, respectively, 
using 4$\times$4$\times$4 ZG supercells. 
It turns out that electron-phonon interactions at finite temperatures result in the closure of the band gap
in SrTiO$_3$, whereas in halide perovskites, electron-phonon interactions cause the opening of the band gap.
Accounting for local disorder in all compounds reduces $\Delta E_g(T)$ by 80--210~meV. 
This is related to the different potential experienced by electrons in the  
disordered network, affecting the electron-phonon matrix elements (see also discussion around Eq.~\eqref{eq.ZG_O_T}). 
Interestingly, for halide perovskites, we observe an almost linear correlation [33~meV/($^\circ$)] 
between the reduction in $\Delta E_g(T)$ due to disorder and the decrease in $\bar{\theta}_{\rm BXB}$.

Figure~\ref{fig5}e shows that experimental values lie within the range of our electron-phonon corrected HSE and PBE0 
band gaps for all disordered compounds. In fact, by adding $\Delta E_g(T)$ to the average PBE0/HSE gap yields 
3.53, 2.08, 1.64, and 1.10~eV for cubic SrTiO$_3$, CsPbBr$_3$, CsPbI$_3$, and CsSnI$_3$, 
respectively, in good agreement with experiments. Our findings here suggest that accurate electronic structure 
calculations of cubic perovskites require the combined corrections due to local disorder, SOC, functionals beyond DFT~\cite{Wiktor2017}, 
and electron-phonon coupling.

In Figs.~\ref{fig5}f,g we show the temperature dependence of the band gap renormalization
evaluated for the high-symmetry (red) and disordered (blue) cubic SrTiO$_3$ and CsPbBr$_3$ using the A-SDM. 
Our calculations for high-symmetry SrTiO$_3$ underestimate experimental data (black) from Ref.~\cite{Kok2015}.
This underestimation is reduced when the disordered network is employed. In fact, 
electron-phonon coupling is strongly modified inducing a correction to the band gap closing of $\sim30$\%. 
This finding together with the computed diffuse scattering patterns further support the presence of 
local disorder in cubic SrTiO$_3$. 
As seen in Fig.~\ref{fig5}g, using the disordered CsPbBr$_3$ also provides an accurate description
of the band gap renormalization, explaining the low variation of the experimental data with temperature~\cite{Sutton2018}. 
Our analysis (see Methods) yields that the low-energy anharmonic optical vibrations dominate electron-phonon coupling 
in locally disordered halide perovskites, contributing 88\% to the band gap renormalization, but 
strongly departing from the simplified picture of a Fr\"ohlich interaction related to harmonic modes.
This finding is consistent with photoluminescense spectra measurements of halide perovskite nanocrystals
which suggest a dominant (negligible) contribution of low-energy optical vibrations (acoustic phonons) to exciton-phonon 
coupling~\cite{Fu2018}. The band gap renormalization calculated using the high-symmetry structure is consistently 300\% larger 
than experiment, fact that further casts doubt on the use of a fully-ordered network in 
first-principles calculations of cubic halide perovskites.
The remarkable success of the cubic polymorphs in describing electron-phonon coupling is explained by 
inspecting the electron lifetimes in halide perovskites~($\sim$4-6~fs)~\cite{Karakus2015,Schlipf2108}
which are much smaller relatively to the period of atomic vibrations ($>$ 200 fs). Hence, anharmonic structural fluctuations 
look essentially static to the electrons which follow the nuclei in their most probable ground state configuration.
We remark that polaronic effects on the band gap renormalization are not included in our calculations. 
Although it is now possible to combine {\it ab initio} polaron distortions~\cite{Hyungjun2023} 
with the A-SDM, such calculations are rather challenging and beyond the scope of this work. We also 
note that corrections to the band gap renormalization coming from hybrid functionals are less than 1~meV (see Methods). 

\begin{figure}[hbt!]
 \begin{center}
\includegraphics[width=0.50\textwidth]{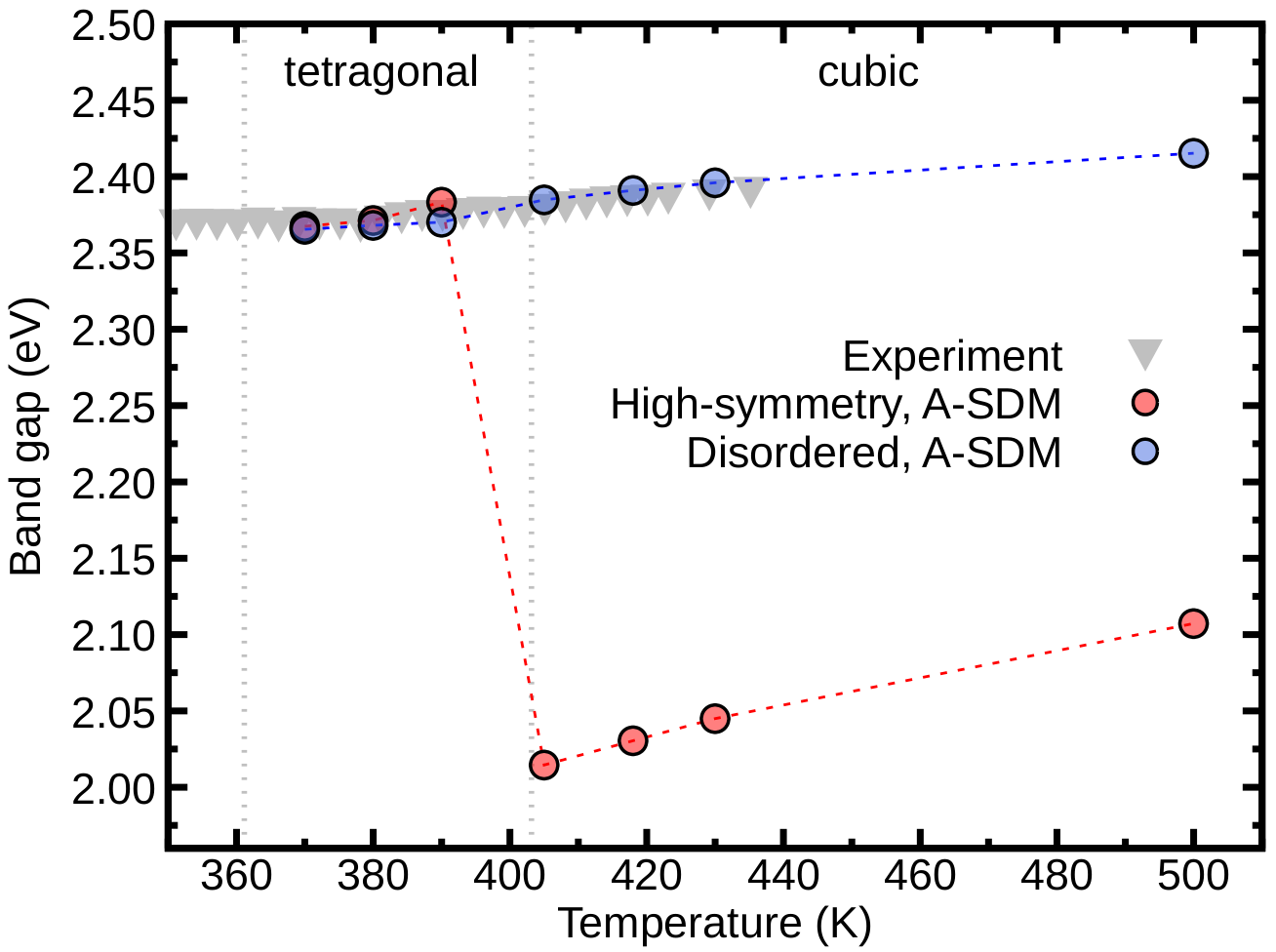}
 \end{center}
\vspace*{-0.4cm}
\caption{\label{fig6}
{\bf Temperature-dependent band gaps of CsPbBr$_3$ across different phases.}
Band gap of CsPbBr$_3$ as a function of temperature calculated for the high-symmetry (red) and disordered (blue)
networks in the tetragonal and cubic phases using the A-SDM.
ZG supercells of size 4$\times$4$\times$4 and 4$\times$2$\times$4, containing 320 atoms each, were 
used for the cubic and tetragonal phases, respectively. ZG displacements for the disordered networks were generated
by taking as a reference the disordered nuclei positions obtained for 2$\times$2$\times$2~supercells. 
DFT-PBEsol data for the high-symmetry and disordered networks are shifted by 1.50 and 1.42~eV, close to the PBE0 
corrections, to match the experimental band gaps at 370~K. The dashed grey lines indicate the phase transition temperatures 
at~361~K and~403~K. Band gap renormalization due to thermal lattice expansion and SOC are accounted for. 
Experimental data (grey) are from Ref.~\cite{Mannino2020}.
}
\end{figure}

In Fig.~\ref{fig6}, we show the band gap variation of CsPbBr$_3$ with temperature calculated within A-SDM
using the cubic and tetragonal phases. Remarkably, acounting for local disorder in our anharmonic electron-phonon coupling 
calculations (blue) yields good agreement with experiment (grey) and captures the smooth variation of the measured band 
gap around the continuous phase transition temperature at 403~K. Instead, using the high-symmetry structures of 
the tetragonal and cubic phases (red), the band gap 
exhibits a spurious abrupt drop of $\sim$0.4~eV, primarily caused by the enforced alignment of the octahedra 
in the cubic phase. A similar issue for the temperature-dependent band gaps computed for the high-symmetry 
CsPbI$_3$ networks has been observed previously~\cite{Ning2022}. 
The continuous change of the band gap from the disordered cubic to the disordered tetragonal phase, achieved here, is consistent with 
a second-order displacive phase transition. We stress that for all computed values we account 
for the same PBE0 corrections to the band gap; that is we apply an identical shift to the temperature-dependent band gaps of both 
the tetragonal and cubic phases. In our calculations for the disordered structures we combined 
A-SDM temperature-dependent IFCs with the disordered networks to generate ZG displacements; examples of the resulting 
phonon spectral functions are shown in Supplementary Figure~17. Notably, using 0~K ground state phonons obtained for 
the disordered structures (Figs.~\ref{fig3}b,c) yields a similar level of agreement (Supplementary Figure~18). 
This demonstrates, essentially, that the vibrational dynamics computed for the locally disordered networks is a reasonable 
approximation to describe anharmonic electron-phonon effects originating from low-energy optical vibrations.

\vspace*{0.1cm}
\noindent {\bf DISCUSSION}
 
Taken all together, our work provides insights on the lattice dynamics and electron-phonon 
couplings in oxide and halide perovskites. 
Given the agreement with measurements of the vibrational-induced 
band gap renormalization of SrTiO$_3$ and CsPbBr$_3$, we expect our approach to be widely used 
for addressing open challenges related to important technological applications of 
halide and oxide perovskites, such as solar cells, light emitting diodes, and thermoelectric devices,
as well as elucidating their unexplored ultrafast spectroscopic properties~\cite{Zhang2023}.
Our results demonstrate that SrTiO$_3$ is fully compatible with a 
static disordered network, while CsPbBr$_3$ is better described within a quasi-static picture that 
captures correctly optical vibrations but not effects arising from ultraslow relaxational rotations 
of the BX$_3$ octahedra. Our findings also confirm that vibrational dynamics in halide perovskites 
deviate from a textbook noninteracting phonon dispersion; 
a picture of strongly coupled vibrations should be considered as a precursor to future calculations 
of perovskites' peculiar transport properties~\cite{Ponce2019,Simoncelli2019}. In the context 
of electron-phonon coupling, the description of tetragonal or cubic perovskites with a locally 
disordered network constitutes the best possible approximation since (i) electron coupling to anharmonic 
optical vibrations is predominant, (ii) dynamical structural fluctuations look essentially static to the short-lived electrons, 
and (iii) electrons mostly see the nuclei as fixed in their disordered ground state. 
This latter point also allows to explain the continuous variation of the band gap around 
phase transitions in halide perovskites, reflecting the subtle rearrangement of atomic positions between
different phases. Some of the important physics of halide perovskites uncovered here 
are intrinsically related to their extraordinary lattice softness~\cite{Ferreira2018}, and, thus, their
remarkable ability to sustain a high degree of disorder. 
At a fundamental level, our study proposes a radically different way of conceptualize
the lattice dynamics in perovskites and sets up a universal framework for accurate simulations of
their carrier mobilities, conductivities, excitonic spectra, non-equilibrium dynamics, and polaron 
physics~\cite{Miyata2017,Miyata2018,Frenzel2023}. 
\\ 

\noindent {\bf METHODS} 

\vspace*{0.2cm}
\noindent {\bf Electronic structure calculations} \\
Electronic structure calculations for SrTiO$_3$, CsPbBr$_3$, CsPbI$_3$, and CsSnI$_3$ 
were performed within density functional theory using
plane waves basis sets as implemented in {\tt Quantum Espresso} (QE)~\cite{QE,QE_2}.
We employed a kinetic energy cutoff of 120~Ry, the Perdew-Burke-Ernzerhof exchange-correlation functional 
revised for solids (PBEsol)~\cite{Perdew_2008}, and optimized norm-conserving Vanderbilt pseudopotentials from 
the PseudoDojo library~\cite{Haman_2013,vanSetten2018}. To account for the effect of SOC 
on the electronic structures we replaced scalar relativistic with fully relativistic pseudopotentials. 
The uniform sampling of the Brillouin zone of the cubic 1$\times$1$\times$1, 2$\times$2$\times$2, 4$\times$4$\times$4,  
and 6$\times$6$\times$6 supercells was performed using 6$\times$6$\times$6, 3$\times$3$\times$3, 
1$\times$1$\times$1, and 1$\times$1$\times$1 ${\bf k}$-grids, respectively.
The only exception is that for 4$\times$4$\times$4 supercells of SrTiO$_3$ we employed a 2$\times$2$\times$2 ${\bf k}$-grid.
Furthermore, for 6$\times$6$\times$6 supercells we reduced the kinetic energy cutoff to 100~Ry. 
Initial calculations of the monomorphous structures were performed in the unit cells of the cubic 
(5 atoms), tetragonal (10 atoms), and orthorhombic (20 atoms) perovskite compounds with 
the nuclei clamped at their Wyckoff positions (space groups: $Pm\bar{3}m$ for all cubic perovskites, 
$I4/mcm$ for tetragonal SrTiO$_3$, $P4/mbm$ for tetragonal CsPbBr$_3$, and $Pbnm$ for orthorhombic CsPbBr$_3$).
The lattice constants of cubic SrTiO$_3$, CsPbBr$_3$, CsPbI$_3$,
and CsSnI$_3$ were fixed to the DFT-PBEsol optimized values of 3.889, 5.874, 6.251, and 6.141~\AA, respectively. 
The lattice constants of the tetragonal and orthorhombic CsPbBr$_3$
were also fixed to the DFT-PBEsol optimized values of ($a=b=5.734$, $c=5.963$~\AA) and ($a = 7.971$, $b = 8.397$,
and $c = 11.640$~\AA). The lattice constants of the tetragonal SrTiO$_3$ were fixed to the experimental lattice 
constants~\cite{Okazaki1973} of ($a=b=3.896$ and $c=3.900$~\AA) since were found to yield better phonon frequency 
renormalizations due to anharmonic effects at finite temperatures~\cite{Zacharias2022}.
The eigenmodes and eigenfrequencies at each phonon wavevector ${\bf q}$ were obtained by evaluating the IFCs 
and corresponding dynamical matrices via the frozen-phonon method~\cite{Kunc_Martin,phonopy}. 
Corrections on the phonon dispersions due to long-range dipole-dipole interactions,
which vary depending on the degree of static-disorder~\cite{Zacharias2022}, were included via the linear response 
approach described in Ref.~\cite{Gonze1997}.

Electron spectral functions of the disordered cubic structures were calculated using the electron band structure
unfolding technique as implemented in the {\tt EPW/ZG} code~\cite{Hyungjun2023,Zacharias_2020}.
We ran calculations with and without SOC (see Supplementary Figures~8 and~9) and sampled the Brillouin zone with 
417 equally-spaced ${\bf k}$-points along the X-R-M-$\Gamma$-R path.  
We remark that when SOC is excluded, local distortions in a 2$\times$2$\times$2 supercell of SrTiO$_3$
lead to an artificial degeneracy splitting of 40 and 60~meV of the triply degenerate valence band top and conduction
band bottom, respectively. Interestingly, our calculations for a 4$\times$4$\times$4 supercell show that
the splitting in the valence band top is eliminated, while in the conduction band bottom is maintained.
This result is consistent with a disordered network that macroscopically
might reflect some of the crystal's symmetries, although local deformations are present.
Inclusion of fully relativistic effects in our calculations for SrTiO$_3$ induces a small band gap change
and spin-orbit splitting of the band edges, as shown in Table~\ref{table.1} and Supplementary Figure~9.
Our fully relativistic calculations for the band gaps of the high-symmetry and locally disordered tetragonal CsPbBr$_3$ within DFT-PBEsol 
yield 0.69 and 0.86~eV, respectively. Ignoring SOC effects our calculations determine 
1.65 and 1.83~eV for the high-symmetry and locally disordered tetragonal CsPbBr$_3$. 
Data calculated for all cubic compounds are provided in Table~\ref{table.1}.

To extract effective masses $m^*$ at the band edges we performed parabolic fits to the electron 
band structure and spectral functions along the specified directions
reported in Supplementary Table~3. The mass enhancement due to disorder, $\lambda$, was obtained from 
 $m^*_{\rm d} = (1+\lambda) \, m^*_{\rm hs}$, where $m^*_{\rm d}$ and $m^*_{\rm hs}$
are the disordered and high-symmetry structure's effective masses. We note that at the proximity of the band edges,
the electron spectral functions give well-defined bands that do not suffer from band broadening.

All calculations employing the Perdew–Burke-Ernzerhof (PBE0)~\cite{pbe0} and Heyd-Scuseria-Ernzerhof 
(HSE06)~\cite{hse06} hybrid functionals 
were performed using the code VASP~\cite{vasp2}. SOC was taken into account, and a 300~eV cut-off energy was set for the 
projector-augmented wave~\cite{paw1}. For the high-symmetry and 2$\times$2$\times$2 supercell disordered structures we employed 
$\Gamma$-centered {\bf k}-grids of 4$\times$4$\times$4 and 2$\times$2$\times$2, respectively.
The HSE06 (PBE0) band gaps with SOC of the high-symmetry and locally disordered tetragonal CsPbBr$_3$ are found to be 
1.20~eV (1.80~eV) and 1.69~eV (2.31~eV), respectively. The corresponding values for all cubic compounds are reported in 
Table~\ref{table.1}. 

\noindent {\bf The locally disordered (polymorphous) network} \\
To explore, initially, the locally disordered network of cubic SrTiO$_3$ we applied three 
different initial sets of displacements 
on the atoms of the high-symmetry structure in a 2$\times$2$\times$2 supercell. Those are: 
(i) ZG displacements along all phonon modes populated at $T=0$~K, 
(ii) ZG displacements along the soft modes populated at $T=0$~K and, 
(iii) random displacements smaller than 0.1~\AA~applied to all atomic coordinates. 
We note that ZG displacements along imaginary soft modes were generated by switching their frequencies to real. 
Each case was followed by a DFT-PBEsol relaxation of the nuclei. 
Although three different ground state disordered geometries were realized, a consistent energy lowering of 8~meV~[f.u.]$^{-1}$ 
relative to the ordered structure was obtained. Starting from random initial displacements 
in a 4$\times$4$\times$4 supercell, the relaxation yields the same 
energy lowering of 8~meV~[f.u.]$^{-1}$ in good agreement with the value of 12~meV~[f.u.]$^{-1}$ reported in Ref.~\cite{Zhao_Zunger2021}.
In Supplementary Figure~10 we demonstrate that the three disordered structures give identical PDFs. 
In Supplementary Table~1 we show that accounting for ground state symmetry-breaking domains in 
4$\times$4$\times$4 supercells yields similar band gap openings with 2$\times$2$\times$2 supercells.

Having demonstrated the equivalence of the three disordered structures, which give the same 
ground state energy and PDF, now we comment on the best choice of 
initial displacements [see (i)-(iii) above].
The most computationally inefficient choice is the use of ZG displacements along all phonon modes, 
bringing the initial configuration well away from its ground state. 
Instead, the relaxation converges much faster when ZG displacements along the soft modes are used which
reproduce the tilting of the octahedra and thus bring the structure closer 
to the bottom of the potential well~\cite{Zacharias2020_FHI}. 
Using random nudges, the efficiency of generating the ground state network 
might vary depending on the system and the amplitude of the initial displacement. 
Therefore, for generating all locally disordered structures, we chose to apply 
special displacements~\cite{Zacharias_2016,Zacharias_2020} along the soft modes  
(computed for the high-symmetry structures) which is the most practical and systematic 
way to achieve ground state optimization. 

The locally disordered (polymorphous) networks of cubic, tetragonal, and orthorhombic phases were explored 
(unless specified otherwise) by employing 2$\times$2$\times$2 supercells containing 40, 80, and 160 atoms, respectively.
To check whether the locally disordered cubic geometries exhibit any residual symmetries, we perform
a symmetry analysis with {\tt pymatgen}~\cite{Ong2013} using a tolerance factor of 0.0001~\AA; we confirm 
that none of the locally disordered cubic structures maintain residual symmetries.

\noindent{\bf Special displacement method}
\\
ZG displacements were generated via the special displacement method~\cite{Zacharias_2016,Zacharias_2020} (SDM) as implemented 
in the {\tt EPW/ZG} code. We used phonons at ${\bf q}$-points commensurate with the supercell size and applied a smooth 
phase evolution of the phonon eigenvectors in reciprocal space. 

Anharmonicity in our calculations was included via the A-SDM  using 2$\times$2$\times$2 supercells
as described in Ref.~\cite{Zacharias2022}. 
Self-consistency in the phonon spectra of each system was achieved using only 3-4 iterations by means of 
a linear mixing scheme. To incorporate the effect of anharmonicity in the phonon-induced band gap
renormalization, we generated ZG displacements in 4$\times$4$\times$4 (all cubic perovskites), 
6$\times$6$\times$6 (cubic SrTiO$_3$), 4$\times$2$\times$4 (tetragonal CsPbBr$_3$), 
and 6$\times$4$\times$4 (tetragonal SrTiO$_3$) supercells employing the IFCs obtained by A-SDM. 
In all calculations of temperature-dependent band gaps reported in Figs.~\ref{fig5}f and g, 
we allowed the lattice to expand according to the measured expansion coefficient~\cite{Ligny1996,Stoumpos2013}.  
Symmetry breaking in ZG configurations led to an artificial degeneracy splitting 
of the band edges of cubic and tetragonal SrTiO$_3$. In this case, the band gap renormalization 
of SrTiO$_3$ at each temperature was evaluated by averaging the energy change of all states participating in the 
formation of the band edges within an energy window of 20~meV. 
To ensure high accuracy and limit the errors arising from artificial degeneracy 
splitting, we also took the average over the band gap renormalization obtained for four different ZG configurations.
In Supplementary Figure~11, we show that the band gap renormalization of cubic SrTiO$_3$ remains 
nearly the same when SDM is combined with 0~K ground state phonons obtained for the disordered structure. 

To identify the contribution of ultraslow acoustic ($E < 2.5$~meV) and low-energy optical 
vibrations ($3.65 < E < 10$~meV) to the band gap renormalization of CsPbBr$_3$, we applied ZG
displacements on the nuclei along only the phonons lying within the associated energy windows. 
We note that using phonons with energies $E > 2.5$~meV and $2.5 < E < 10$~meV we obtain a similar 
band gap renormalization (within 25~meV) which demonstrates, essentially, that high-energy optical 
vibrations ($E > 10$~meV) do not play an important role in the electron-phonon gap renormalization of 
halide perovskites.

All A-SDM calculations for determining the phonon-induced band gap renormalization were performed 
at the DFT-PBEsol level. Corrections to the band gap renormalization arizing from hybrid functionals were found to be 
negligible. In particular, our calculations in 2$\times$2$\times$2 ZG supercells of disordered CsPbBr$_3$ yield a ZPR 
of 29.4, 29.0, and 29.9~meV for DFT-PBEsol, HSE, and PBE0 functionals, respectively.

\noindent{\bf Phonon unfolding} 
\\
For systems undergoing static symmetry breaking due to lattice distortion coming, e.g., from defects, atomic disorder, 
or a charge density wave, a supercell is required to compute the phonons. In this 
case, the crystal's symmetry operations (translations and rotations) are no longer applicable 
and all atoms in the supercell need to be displaced for calculating the dynamical matrix and, hence, the 
renormalized phonon frequencies $\omega_{{\bf Q} \mu}$, where ${\bf Q}$ and $\mu$ are the phonon wavevector 
and band indices. To illustrate the effect of lattice 
distortion in the phonons, a common practise is to employ phonon unfolding and evaluate 
the momentum-resolved spectral function given by~\cite{Allen_2013}:
\begin{equation}\label{eq.sprtl_fn_T}
  A_{\bf q}(\omega) =  \sum_{{\bf Q} \mu} P_{{\bf Q} \mu,{\bf q}} \,\delta(\omega - \omega_{{\bf Q} \mu}). 
 \end{equation}
Here ${\bf q}$ denotes a wavevector in the Brillouin zone of the unit cell and 
$P_{{\bf Q} \mu,{\bf q}}$ represents the spectral weights which are evaluated in the 
spectral representation of the single-particle Green's function as~\cite{Zheng2016}: 
\begin{eqnarray}\label{eq.Ovmat4}
 P_{{\bf Q} \mu,{\bf q}} = \frac{1}{ N_{\bf g} } \frac{\Omega}{\tilde{\Omega}} \sum_{ \alpha j} \bigg| \sum_{\kappa} 
                   \tilde{e}_{\alpha \kappa, \mu}({\bf q}) {\rm e}^{{\rm i} ({\bf q} + {\bf g}_{j}) \cdot \tilde{\boldsymbol{ \tau}}_{\kappa}} \bigg|^2, 
\end{eqnarray}
where $j$ is an index for the reciprocal lattice vectors ${\bf g}$ of the unit's cell Brillouin zone, $\alpha$ denotes a 
Cartesian direction, and $\kappa$ is the atom index.
The symbol $\sim$ indicates quantities calculated using the disordered structure.
$N_{\bf g}$ acts as a normalization factor representing the total number of 
reciprocal lattice vectors entering the summation. 
The spectral weight can be understood, essentially, as the projection of the phonon eigenvector 
$\tilde{e}_{\alpha \kappa, \mu}({\bf Q})$ on the phonon eigenvectors $e_{\alpha\kappa, \nu}({\bf q})$ computed in the unit cell, 
given that ${\bf Q}$ unfolds into ${\bf q}$ via ${\bf Q} = {\bf q} + {\bf g}_j - {\bf G}$, where ${\bf G}$ is 
a reciprocal lattice vector of the distorted structure.

To generate vibrational spectral functions we employed Eq.~\eqref{eq.Ovmat4} 
and 417, 382, and 399 equally-spaced ${\bf q}$-points along the X-R-M-$\Gamma$-R 
(cubic),  X-A-M-$\Gamma$-A (tetragonal), and X-R-S-$\Gamma$-R (orthorhombic) paths. 
Convergence of the spectral weights was ensured by using a 10$\times$10$\times$10
${\bf g}$-grid of reciprocal lattice vectors. 
In Supplementary Figures~12 and~13 we demonstrate our implementation of phonon unfolding by comparing 
phonon spectral functions computed for 2$\times$2$\times$2 and 4$\times$4$\times$4 supercells of SrTiO$_3$. 
Our implementation of phonon unfolding is available in the {\tt EPW/ZG} tree.
In Supplementary Figure~14, we also show that the vibrational spectrum of CsPbBr$_3$ remains almost identical 
when two different ground state disordered geometries are considered. 

We note that overdamped unfolded phonon spectra of halide perovskites 
has also been revealed by analysis of velocity autocorrelation functions obtained by molecular dynamics 
simulations~\cite{Lahnsteiner2022}. 

\noindent {\bf Diffuse scattering} \\
All-phonon diffuse scattering maps were calculated within the Laval-Born-James (LBJ) theory using {\tt disca.x} 
of the {\tt EPW/ZG} code~\cite{Zacharias2021_allph,Zacharias2021_multiph}. 
The merit of the LBJ theory is that inelastic scattering arising from one-phonon and 
multiphonon processes is accounted for on the same basis. 
The Debye-Waller and phononic factors entering LBJ theory [Eq.~(1) of Ref.~\cite{Zacharias2021_allph}]
were evaluated for a $16\!\times\!16\!\times\!16$ ${\bf q}$-grid. The phonon eigenmodes and frequencies
were obtained by means of Fourier interpolation of the dynamical matrices computed for 2$\times$2$\times$2
supercells using either the A-SDM or the disordered network. A $16\!\times\!16\!\times\!1$ 
uniform ${\bf Q}$-grid (scattering wavevectors) per Brillouin zone was used to calculate the 
phonon-induced scattering intensity in the reciprocal lattice planes 
perpendicular to one Cartesian axis. The atomic scattering
amplitudes were determined as a sum of Gaussians with the parameters taken 
from Ref.~\cite{Peng_book}. The diffuse scattering maps of CsPbBr$_3$ for ultraslow acoustic and 
low-energy phonon dynamics were determined by excluding the modes outside the associated energy windows. 
In addition to the scattering maps of cubic SrTiO$_3$ presented in Fig.~\ref{fig4}, we also 
calculated diffuse scattering in the $(Q_x, Q_y, 0)$ and $(Q_x, Q_y, 1)$ planes and 
found qualitative agreement with measurements 
of Ref.~\cite{Kopeck2012} (Supplementary Figures~5 and~6).
In Supplementary Figures~15 and~16, we show the decomposition of the all-phonon scattering in the $(Q_x, Q_y, 0)$
planes of cubic SrTiO$_3$ and CsPbBr$_3$ into one-phonon and multiphonon processes for a large range 
of scattering wavevectors.

{\bf DATA AVAILABILITY} \\
The calculations (input and output files) employed for this study are available via the NOMAD archive
[{\color{blue}http://doi.org/10.17172/NOMAD/2023.07.11-1} for anharmonic electron-phonon coupling calculations 
and {\color{blue}http://doi.org/10.17172/NOMAD/2023.05.13-1} for anharmonic phonon calculations],
or upon request from the corresponding author. 

{\bf CODE AVAILABILITY} \\
{QUANTUM ESPRESSO} is available under GNU General Public Licence from the {QUANTUM ESPRESSO} 
web site ({\color{blue}https://www.quantum-espresso.org/}).
The ZG module of EPW employed for the treatment of local disorder, anharmonicity, and generation of anharmonic self-consistent 
special displacements is also available at GitLab ({\color{blue}https://gitlab.com/epw-code/q-e/tree/ZG}).

{\bf ACKNOWLEDGEMENTS} \\
We thank M. Kopeck{\'{y}} and J. F{\'{a}}bry for kindly sharing experimental data of SrTiO$_3$ 
diffuse scattering maps. We also thank D. R. Ceratti for graciously providing data on 
temperature-dependent band gaps of CsPbBr$_3$ single crystals.
 M.Z. acknowledges funding from the European Union’s Horizon 2020 research and innovation programme
under the Marie Skłodowska-Curie Grant Agreement No. 899546.
This research was also funded by the European Union (project ULTRA-2DPK / HORIZON-MSCA-2022-PF-01 /
Grant Agreement No. 101106654). Views and opinions expressed are however those of the authors only and do not necessarily 
reflect those of the European Union or the European Commission. Neither the European Union nor the 
granting authority can be held responsible for them.
J.E. acknowledges financial support from the Institut Universitaire de France.
F.G. was supported by the National Science Foundation under CSSI Grant No. 2103991 and DMREF Grant No. 2119555.
The work at institute FOTON and ISCR was supported by the European Union’s Horizon 2020 research and 
innovation program under grant agreement 861985 (PeroCUBE) and grant agreement 899141 (PoLLoC). 
We acknowledge that the results of this research have been achieved using the DECI resource
Prometheus at CYFRONET in Poland [https://www.cyfronet.pl/] with support from the PRACE aisbl
and HPC resources from the Texas Advanced Computing Center (TACC) at The University of Texas at Austin
[http://www.tacc.utexas.edu].

\newpage 

\bibliographystyle{naturemag}
\bibliography{references}

\newpage

\end{document}